\documentclass[journal]{IEEEtran}
\usepackage{url}
\usepackage[utf8]{inputenc}
\usepackage{xcolor}
\usepackage{amsmath}
\usepackage{amssymb}

\usepackage[acronyms,nonumberlist,nopostdot,nomain,nogroupskip]{glossaries}
\usepackage{tablefootnote}
\usepackage{booktabs}
\usepackage{tabularx}
\usepackage{tikz}
\usepackage{pgfplots}
\pgfplotsset{compat=newest} 
\pgfplotsset{plot coordinates/math parser=false} 
\newlength\fheight
\newlength\fwidth
\usetikzlibrary{plotmarks,patterns,decorations.pathreplacing,backgrounds,calc,arrows,arrows.meta,spy,matrix}
\usepgfplotslibrary{patchplots,groupplots}
\usepackage{tikzscale}
\usepackage{hyperref}

\usepackage{multirow}
\usepackage{tkz-kiviat}

\usepackage[font=scriptsize]{subcaption}
\usepackage[font=footnotesize]{caption}

\usepackage{mathtools}

\usepackage{dblfloatfix}    
\usepackage{colortbl}

\newacronym{3gpp}{3GPP}{3rd Generation Partnership Project}
\newacronym{adc}{ADC}{Analog to Digital Converter}
\newacronym{5g}{5G}{5th generation}
\newacronym{aimd}{AIMD}{Additive Increase Multiplicative Decrease}
\newacronym{am}{AM}{Acknowledged Mode}
\newacronym{amc}{AMC}{Adaptive Modulation and Coding}
\newacronym{aqm}{AQM}{Active Queue Management}
\newacronym{awgn}{AGWN}{Additive White Gaussian Noise}
\newacronym{balia}{BALIA}{Balanced Link Adaptation}
\newacronym{bdp}{BDP}{Bandwidth-Delay Product}
\newacronym{bf}{BF}{Beamforming}
\newacronym{cc}{CC}{Congestion Control}
\newacronym{cdf}{CDF}{Cumulative Distribution Function}
\newacronym{cn}{CN}{Core Network}
\newacronym{cqi}{CQI}{Channel Quality Information}
\newacronym{cp}{CP}{Control Plane}
\newacronym{csirs}{CSI-RS}{Channel State Information - Reference Signal}
\newacronym{dc}{DC}{Dual Connectivity}
\newacronym{dce}{DCE}{Direct Code Execution}
\newacronym{dci}{DCI}{Downlink Control Information}
\newacronym{dl}{DL}{Downlink}
\newacronym{dmr}{DMR}{Deadline Miss Ratio}
\newacronym{dmrs}{DMRS}{DeModulation Reference Signal}
\newacronym{bsm}{BSM}{Basic Safety Message}
\newacronym{cam}{CAM}{Cooperative Awareness Message}
\newacronym{e2e}{E2E}{End-to-End}
\newacronym{ecn}{ECN}{Explicit Congestion Notification}
\newacronym{edf}{EDF}{Earliest Deadline First}
\newacronym{enb}{eNB}{evolved Node Base}
\newacronym{epc}{EPC}{Evolved Packet Core}
\newacronym{es}{ES}{Edge Server}
\newacronym{fdma}{FDMA}{Frequency Division Multiple Access}
\newacronym{fdd}{FDD}{Frequency Division Duplexing}
\newacronym[firstplural=Radio Access Technologies (RATs)]{rat}{RAT}{Radio Access Technology}
\newacronym{fs}{FS}{Fast Switching}
\newacronym{ftp}{FTP}{File Transfer Protocol}
\newacronym{gnb}{gNB}{Next Generation Node Base}
\newacronym{harq}{HARQ}{Hybrid Automatic Repeat reQuest}
\newacronym{hetnet}{HetNet}{Heterogeneous Network}
\newacronym{hh}{HH}{Hard Handover}
\newacronym{hol}{HOL}{Head-of-Line}
\newacronym{ia}{IA}{Initial Access}
\newacronym{imt}{IMT}{International Mobile Telecommunication}
\newacronym{iot}{IoT}{Internet of Things}
\newacronym{los}{LOS}{Line of Sight}
\newacronym{lte}{LTE}{Long Term Evolution}
\newacronym{m2m}{M2M}{Machine to Machine}
\newacronym{mac}{MAC}{Medium Access Control}
\newacronym{mc}{MC}{Multi-Connectivity}
\newacronym{mcs}{MCS}{Modulation and Coding Scheme}
\newacronym{mec}{MEC}{Mobile Edge Cloud}
\newacronym{mi}{MI}{Mutual Information}
\newacronym{mimo}{MIMO}{Multiple Input, Multiple Output}
\newacronym{mmwave}{mmWave}{millimeter wave}
\newacronym{mptcp}{MPTCP}{Multipath TCP}
\newacronym{mr}{MR}{Maximum Rate}
\newacronym{mss}{MSS}{Maximum Segment Size}
\newacronym{mtd}{MTD}{Machine-Type Device}
\newacronym{mtu}{MTU}{Maximum Transmission Unit}
\newacronym{nfv}{NFV}{Network Function Virtualization}
\newacronym{nlos}{NLOS}{Non Line of Sight}
\newacronym{nr}{NR}{NR}
\newacronym{ofdm}{OFDM}{Orthogonal Frequency Division Multiplexing}
\newacronym{pdcch}{PDCCH}{Physical Downlonk Control Channel}
\newacronym{pdcp}{PDCP}{Packet Data Convergence Protocol}
\newacronym{pdsch}{PDSCH}{Physical Downlink Shared Channel}
\newacronym{pdu}{PDU}{Packet Data Unit}
\newacronym{pf}{PF}{Proportional Fair}
\newacronym{pgw}{PGW}{Packet Gateway}
\newacronym{phy}{PHY}{Physical}
\newacronym{pbch}{PBCH}{Physical Broadcast Channel}
\newacronym[plural=\gls{mme}s,firstplural=Mobility Management Entities (MMEs)]{mme}{MME}{Mobility Management Entity}
\newacronym{prb}{PRB}{Physical Resource Block}
\newacronym{pss}{PSS}{Primary Synchronization Signal}
\newacronym{pucch}{PUCCH}{Physical Uplink Control Channel}
\newacronym{pusch}{PUSCH}{Physical Uplink Shared Channel}
\newacronym{rach}{RACH}{Random Access Channel}
\newacronym{ran}{RAN}{Radio Access Network}
\newacronym{red}{RED}{Random Early Detection}
\newacronym{rf}{RF}{Radio Frequency}
\newacronym{rlc}{RLC}{Radio Link Control}
\newacronym{rlf}{RLF}{Radio Link Failure}
\newacronym{rrc}{RRC}{Radio Resource Control}
\newacronym{rrm}{RRM}{Radio Resource Management}
\newacronym{rr}{RR}{Round Robin}
\newacronym{rs}{RS}{Remote Server}
\newacronym{rsrp}{RSRP}{Reference Signal Received Power}
\newacronym{rss}{RSS}{Received Signal Strength}
\newacronym{rtt}{RTT}{Round Trip Time}
\newacronym{rw}{RW}{Receive Window}
\newacronym{rx}{RX}{Receiver}
\newacronym{sa}{SA}{standalone}
\newacronym{sack}{SACK}{Selective Acknowledgment}
\newacronym{sap}{SAP}{Service Access Point}
\newacronym{sch}{SCH}{Secondary Cell Handover}
\newacronym{scoot}{SCOOT}{Split Cycle Offset Optimization Technique}
\newacronym{sdma}{SDMA}{Spatial Division Multiple Access}
\newacronym{sinr}{SINR}{Signal to Interference plus Noise Ratio}
\newacronym{sm}{SM}{Saturation Mode}
\newacronym{snr}{SNR}{Signal to Noise Ratio}
\newacronym{son}{SON}{Self-Organizing Network}
\newacronym{ss}{SS}{Synchronization Signal}
\newacronym{srs}{SRS}{Sounding Reference Signal}
\newacronym{sss}{SSS}{Secondary Synchronization Signal}
\newacronym{tb}{TB}{Transport Block}
\newacronym{tcp}{TCP}{Transmission Control Protocol}
\newacronym{tdd}{TDD}{Time Division Duplexing}
\newacronym{tdma}{TDMA}{Time Division Multiple Access}
\newacronym{tfl}{TfL}{Transport for London}
\newacronym{tm}{TM}{Transparent Mode}
\newacronym{trp}{TRP}{Transmitter Receiver Pair}
\newacronym{tti}{TTI}{Transmission Time Interval}
\newacronym{ttt}{TTT}{Time-to-Trigger}
\newacronym{tx}{TX}{Transmitter}
\newacronym{ue}{UE}{User Equipment}
\newacronym{ul}{UL}{Uplink}
\newacronym{uml}{UML}{Unified Modeling Language}
\newacronym{um}{UM}{Unacknowledged Mode}
\newacronym{utc}{UTC}{Urban Traffic Control}
\newacronym{vm}{VM}{Virtual Machine}
\newacronym{rsrq}{RSRQ}{Reference Signal Received Quality}
\newacronym{rssi}{RSSI}{Received Signal Strength Indicator}
\newacronym{crs}{CRS}{Cell Reference Signal}
\newacronym{v2v}{V2V}{Vehicle-to-Vehicle}
\newacronym{v2i}{V2I}{Vehicle-to-Infrastructure}
\newacronym{v2n}{V2N}{Vehicle-to-Network}
\newacronym{v2x}{V2X}{Vehicle-to-Everything}
\newacronym{vn}{VN}{Vehicular Node}
\newacronym{dsrc}{DSRC}{Dedicated Short Range Communication}
\newacronym{ci}{CI}{context information}
\newacronym{voi}{VoI}{value of information}
\newacronym{aoi}{AoI}{age of information}
\newacronym{gps}{GPS}{Global Positioning System}
\newacronym{qos}{QoS}{Quality of Service}
\newacronym{qoe}{QoE}{Quality of Experience}
\newacronym{ml}{ML}{Machine Learning}
\newacronym{ahp}{AHP}{Analytic Hierarchy Process}
\newacronym{lidar}{LIDAR}{Light Detection and Ranging}
\newacronym{c-its}{C-ITS}{Cooperative and Intelligent Transportation System}
\tikzstyle{startstop} = [rectangle, rounded corners, minimum width=2cm, minimum height=0.5cm,text centered, draw=black]
\tikzstyle{io} = [trapezium, trapezium left angle=70, trapezium right angle=110, minimum width=3cm, minimum height=1cm, text centered, draw=black]
\tikzstyle{process} = [rectangle, minimum width=2cm, minimum height=0.5cm, text centered, draw=black, alignb=center]
\tikzstyle{decision} = [ellipse, minimum width=2cm, minimum height=1cm, text centered, draw=black]
\tikzstyle{arrow} = [thick,<->,>=stealth]
\tikzstyle{line} = [thick,>=stealth]
\tikzstyle{darrow} = [thick,<->,>=stealth,dashed]
\tikzstyle{sarrow} = [thick,->,>=stealth]
\tikzstyle{larrow} = [line width=0.1mm,dashdotted,->,>=stealth]

\makeatletter
\def\grd@save@target#1{%
  \def\grd@target{#1}}
\def\grd@save@start#1{%
  \def\grd@start{#1}}
\tikzset{
  grid with coordinates/.style={
    to path={%
      \pgfextra{%
        \edef\grd@@target{(\tikztotarget)}%
        \tikz@scan@one@point\grd@save@target\grd@@target\relax
        \edef\grd@@start{(\tikztostart)}%
        \tikz@scan@one@point\grd@save@start\grd@@start\relax
        \draw[minor help lines] (\tikztostart) grid (\tikztotarget);
        \draw[major help lines] (\tikztostart) grid (\tikztotarget);
        \grd@start
        \pgfmathsetmacro{\grd@xa}{\the\pgf@x/1cm}
        \pgfmathsetmacro{\grd@ya}{\the\pgf@y/1cm}
        \grd@target
        \pgfmathsetmacro{\grd@xb}{\the\pgf@x/1cm}
        \pgfmathsetmacro{\grd@yb}{\the\pgf@y/1cm}
        \pgfmathsetmacro{\grd@xc}{\grd@xa + \pgfkeysvalueof{/tikz/grid with coordinates/major step x}}
        \pgfmathsetmacro{\grd@yc}{\grd@ya + \pgfkeysvalueof{/tikz/grid with coordinates/major step y}}
        \foreach \x in {\grd@xa,\grd@xc,...,\grd@xb}
        \node[anchor=north] at (\x,\grd@ya) {\pgfmathprintnumber{\x}};
        \foreach \y in {\grd@ya,\grd@yc,...,\grd@yb}
        \node[anchor=east] at (\grd@xa,\y) {\pgfmathprintnumber{\y}};
      }
    }
  },
  minor help lines/.style={
    help lines,
    gray,
    line cap =round,
    xstep=\pgfkeysvalueof{/tikz/grid with coordinates/minor step x},
    ystep=\pgfkeysvalueof{/tikz/grid with coordinates/minor step y}
  },
  major help lines/.style={
    help lines,
    line cap =round,
    line width=\pgfkeysvalueof{/tikz/grid with coordinates/major line width},
    xstep=\pgfkeysvalueof{/tikz/grid with coordinates/major step x},
    ystep=\pgfkeysvalueof{/tikz/grid with coordinates/major step y}
  },
  grid with coordinates/.cd,
  minor step x/.initial=.5,
  minor step y/.initial=.2,
  major step x/.initial=1,
  major step y/.initial=1,
  major line width/.initial=1pt,
}
\makeatother
\usepackage{makecell}
\usepackage{diagbox}
\usepackage{tikz-qtree}
\usetikzlibrary{trees} 
\newcolumntype{b}{X}
\newcolumntype{s}{>{\hsize=.01\hsize}X}

\makeglossaries

\begin{document}

\title{\fontsize{23}{25}\selectfont Investigating  Value of Information \\  in   Future  Vehicular  Communications }

\author{\IEEEauthorblockN{ Marco Giordani$^{\circ }$,  Andrea Zanella$^{\circ }$, Takamasa Higuchi$^{\dagger }$, Onur Altintas$^{\dagger}$, Michele Zorzi$^{\circ }$}
\IEEEauthorblockA{\\
$^{\circ }$Department of Information Engineering (DEI), University of Padova, Italy \\
$^{\dagger}$InfoTech Labs, Toyota Motor North America, Inc.\\
Email: {\{giordani,zanella,zorzi\}@dei.unipd.it}, {\{takamasa.higuchi,onur.altintas\}@toyota.com }\vspace{-0.99cm}}}

\maketitle

\begin{abstract}
The next generations of vehicles are expected to be equipped with sophisticated sensors to support advanced automotive services. The large volume of data generated by such  applications will likely put a strain on the vehicular communication technologies, which may be unable to guarantee the required quality of service. In this scenario, it is fundamental to assess the \emph{value of  information} (VoI) provided by each data source, to prioritize the transmissions that have greatest importance for the target applications. 
In this paper, we characterize VoI in future vehicular networks, and investigate efficient data dissemination methods to tackle capacity issues.
Through a simulation study, we show  how analytic hierarchy multicriteria decision processes can be exploited to  determine the value of sensory observations as a function of space, time, and~quality criteria.
\end{abstract}


\vspace{-0.63cm}
\section{Introduction}
\label{sec:introduction}
In recent years,   the automotive industry has  evolved towards \glspl{c-its} to offer safer traveling and improved traffic management. 
Connectivity among vehicles, i.e., V2X, has also emerged as a means to enable advanced automated driving applications whose unprecedentedly stringent demands, e.g., in terms of data rate, reliability and latency~\cite{3GPP_22186}, may however saturate the capacity of traditional technologies for vehicular communications~\cite{giordani2018feasibility,giordaniperformance2018}.
The scientific community is  working towards the development of new radio systems, e.g., operating at \glspl{mmwave}, that may cope with these challenges.
This potential is however hindered by the harsh propagation characteristics of the above-6 GHz bands~\cite{magazine2016_Heath,MOCAST_2017}.

We argue that even a significant increase in the channel capacity may not be sufficient to satisfy the boldest \gls{qos} requirements of future automotive applications, in particular in scenarios with multiple active services  requiring different degrees of automation. 
In this context, it becomes fundamental to limit the amount of information that can be broadcast over bandwidth-constrained communication channels.
One approach is to set a bound on the \emph{\gls{aoi}}~\cite{kaul2012real}, by making vehicles broadcast awareness messages that are never older than the inter-transmission period. 
Another approach is to discriminate the \emph{\gls{voi}}~\cite{howard1966information}, to use the limited transmission resources in a way that  maximizes the utility for the target~applications.


Traditionally,  \gls{voi} has been studied under an economic perspective to support data management and decision making~\cite{wang1996beyond}.
These strategies, however, typically assume that information is consumed by humans, therefore they might not be applicable in the context of \glspl{c-its}.

The \gls{voi} theory has also been applied to underwater systems to  decide  how much information  to transmit through resource-constrained networks~\cite{basagni2014maximizing}.
Such techniques, however, present many limitations in a V2X context, due to the completely different characteristics of acoustic propagation compared to the higher-frequency higher-bandwidth vehicular~channel.

\gls{voi} has also been investigated in a military context to prioritize the information  to be disseminated to or gathered from  soldiers in a battlefield environment~\cite{suri2015exploring}. 
These strategies, however, do not account for cases where the information sources are not directly under the  users' control (e.g., for data automatically generated by a vehicle's sensors).

Recently, \gls{voi}-aware methods have  been proposed for sensor network  applications to provide data that fulfills the needs of users under resource, monetary and latency constraints~\cite{bisdikian2013quality}.
Sensor network optimization, however, has been mostly associated to traditional \gls{qos} paradigms, e.g., to minimize  power consumption~\cite{fasolo2007network}.
In the vehicular context, instead,  sensory information  should  be valued based on  the utility provided to the final receiver  in a specific scenario.
Moreover,  VoI is expected to change much more significantly than considering static sensors deployed, e.g., in smart~cities.

 \begin{figure*}[t!]
 \centering
 \setlength{\belowcaptionskip}{-0.33cm}
 \includegraphics[width=0.8\textwidth]{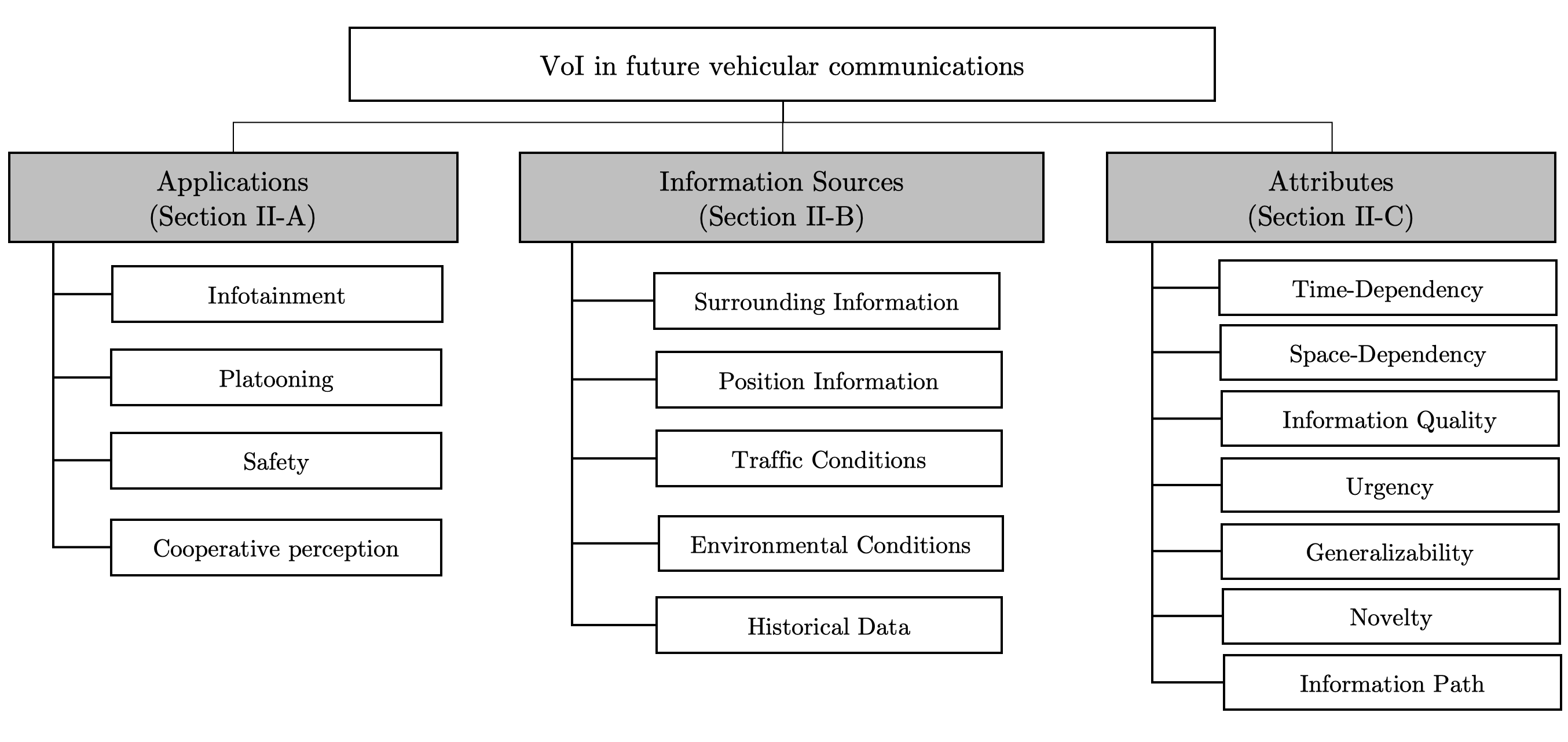}
 \vspace{-0.33cm}
 \caption{Classifications of elements taking part in the definition of \gls{voi} in future vehicular communications. }
    \label{fig:fig_taxonomy}
 \end{figure*}



Most existing value assessment approaches are based on the so-called urgency-aware prioritization mechanisms, which balance  delivery of non-critical information and timely dissemination of high-priority data.
On the other hand, considering the transient nature of V2X topology,  the  VoI tends to decrease in time and may become outdated quite rapidly, thereby requiring a dynamic adjustment of the value which should account for specific attributes, e.g., \emph{quality, consistency,~timeliness}.
Following this rationale, in this paper we characterize \gls{voi} in vehicular networks, and investigate its applicability to promote efficient information distribution, preventing the overload of transmission links.

Among the original contributions of this work, in Sec.~\ref{sec:voi_parameter_preliminaries} we provide a taxonomy of information sources in vehicular networks and  identify the  attribute categories that should be accounted for in the \gls{voi} assessment process.
Then,  in Sec.~\ref{sec:voi_assignment_techniques} we list the benefits and the limitations of some of the most relevant  methods  proposed in the literature to assess the \gls{voi}, which help enable  information dissemination when considering the requirements of future vehicular communication systems. 
In Sec.~\ref{sec:voi_assessment_the_analytic_hierarchy_process_approach} we finally show, through an illustrative example, how to exploit analytic hierarchy multicriteria decision processes to qualitatively assign \gls{voi} and rank each piece of information in order to  best support the needs of users  in  resource-constrained networks.
The results prove that
the value of  information may increase or decrease significantly according to spatial, temporal and accuracy criteria.

\section{VoI Assignment: Preliminaries for V2X} 
\label{sec:voi_parameter_preliminaries}

This section provides an overview of the requirements of some automotive applications, the types of information they can use, and the attributes to be considered to assess their value, as schematically represented in Fig.~\ref{fig:fig_taxonomy}.

\vspace{-0.2cm}
\subsection{Applications Taxonomy} 
\label{sub:application_requirements}
We focus on four representative application domains whose requirements have been  outlined by the 3GPP in~\cite{3GPP_22186}.

\emph{a) Infotainment} refers to a set of services  that deliver a combination of information and entertainment, e.g., video streaming and on-line gaming. These applications have stringent demands in terms of high throughput and low latency,  and may require the dynamic maintenance of multicast~communication.



\emph{b) Safety} 
applications  require the exchange of short messages (up to a few hundreds of bytes), but with strict latency  (around 10~ms) and reliability  ($>$ 99.99\%)~constraints. 

\emph{c) Cooperative perception (advanced driving)} 
enables semi- or fully-automated driving through persistent dissemination of perception data.
These operations may require  high-throughput connections (up to hundreds of Mbps), due to the detailed nature of the shared contents, while some latency could be tolerated depending on the degree of automation.

\emph{d) Platooning} 
refers to the services (e.g., cooperative adaptive cruise control) that make it possible for a group of vehicles that follow the same trajectory to travel in close proximity to one another at highway speeds. 
A significant amount of information needs to be shared via V2X communications. In addition to the strict latency requirement, the connection reliability and stability are also very critical.


\vspace{-0.2cm}
\subsection{Information Sources Taxonomy} 
\label{sub:ci_taxonomy}
Here we list some types of information that can be collected and exchanged by vehicles to support automotive services. 

\paragraph{Surrounding Information} 
\label{par:surrounding_information}
The awareness of the surrounding environment is the basis of a wide set of services. Such information is typically provided by basic on-board sensors, e.g.,
radars and sonars, which enable detection and localization of surrounding objects.
More sophisticated instrumentation, e.g., cameras and \gls{lidar} sensors, can be used for road signs recognition and classification, and generate high-bitrate raw data flows (from 100 Mbps to 700 Mbps, depending on the image quality)~\cite{magazine2016_Heath}. 
 
 \begin{figure*}[t!]
 \centering
 \setlength{\belowcaptionskip}{-0.5cm}
 \includegraphics[width=0.85\textwidth]{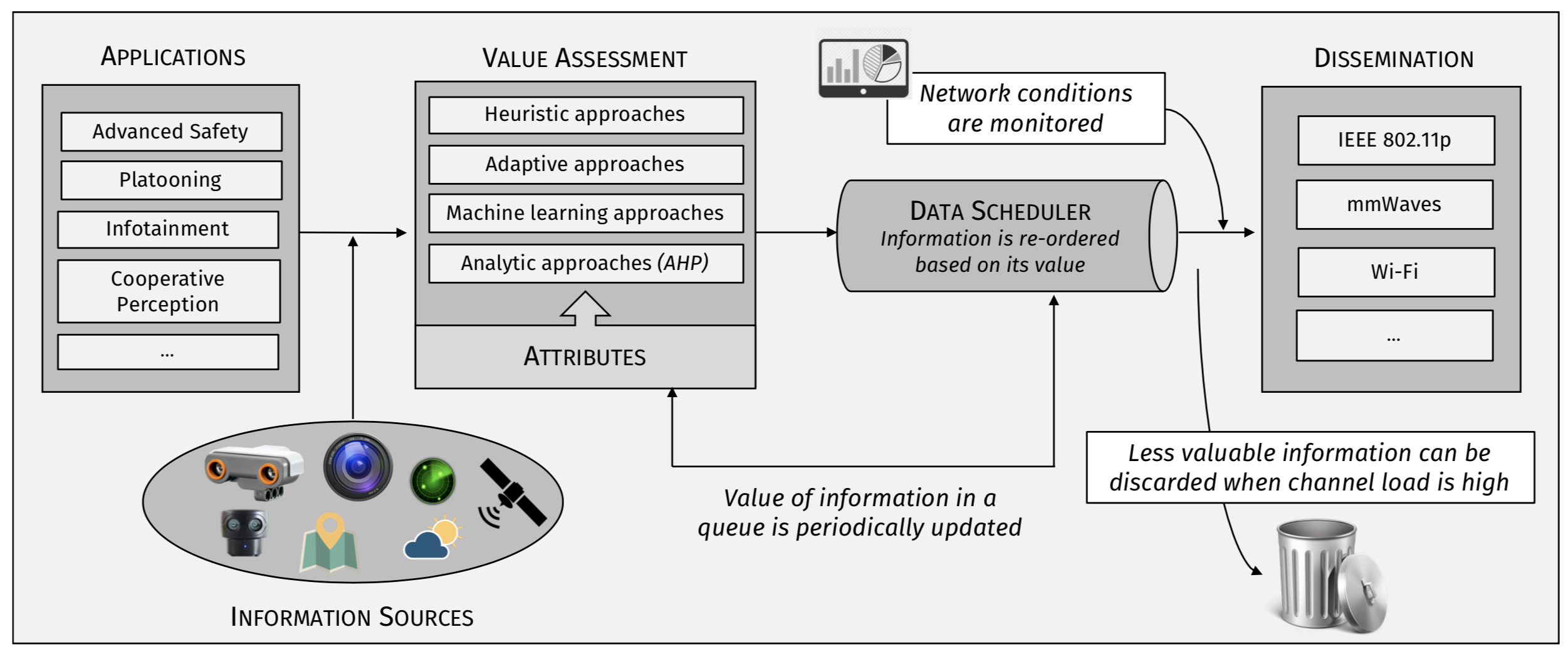}
 \caption{\footnotesize  Block diagram of the  \gls{voi} assessment process.  }
    \label{fig:scheme}
 \end{figure*}


\paragraph{Position Information} 
 \gls{gps} offers global time synchronization and absolute (though not always accurate) positioning, although other localization techniques, e.g., based on dead reckoning or image processing, may be required to improve positioning accuracy. Through \glspl{bsm}, vehicles can also determine their mutual speeds and accelerations. 

\paragraph{Traffic Conditions and Prediction} 
Real-time traffic information can be obtained collecting vehicles' localization measurements and re-broadcasted by network infrastructures.
Traffic information may also be complemented with side information, e.g., the presence of sensitive locations (schools, hospitals) or temporary
events (city marathons, social/political events), which may have an impact on the traffic in the area. 

 \paragraph{Environmental Conditions} 
Weather conditions, including rain, snow,  fog, dust, ice and black ice, can be provided by national weather-alerting systems and made accessible through network infrastructures. The corresponding information flow is light, with loose reliability and latency~constraints. 

  \paragraph{Historical Data} 
  Past observations can be turned into experience. 
  As a vehicle is able to recognize a specific profile (e.g., a driver, a place, a road), it can access its saved historical data and exploit this information, e.g., to adapt its driving decisions. 
  Such historical data may not be available from the vehicles currently on the road, but can be stored in infrastructure or cloud servers and downloaded when required. 
  Data traffic flow is expected to be low, with low latency~requirements.



\vspace{-0.1cm}
\subsection{Attributes Taxonomy} 
\label{sub:attributes_taxonomy}
The value of the information can be assessed based on multiple attributes, categorized in the following. 




\paragraph{Time Dependency} 
Typically, the \gls{voi} in vehicular networks is affected by obsolescence, so that the value  decays over time in a way that 
is highly application-dependent. 
For example, safety messages are extremely delay sensitive, while infotainment data is more time-resilient, since buffering and error concealment techniques may maintain good service quality even in the presence of communication~gaps. 

\paragraph{Space Dependency} 
Analogously, we can define an application-dependent spatial horizon over which a piece of information is valuable for the potential receiver(s). 
For example, for collision-warning applications, sensory data generated by close-by vehicles are more valuable than those coming from farther nodes, while the situation can be the opposite for some perception applications. 

\paragraph{Information Quality} 
\gls{voi} may  depend on the \emph{quality} of  the data, which may be assessed in terms of accuracy (e.g., for \gls{gps} coordinates), resolution (e.g., video), and variance. The importance of quality attributes depends on the target application, e.g., high-definition \gls{lidar} images may be of little value for infotainment services, but very precious for safety services instead.

\paragraph{Urgency} 
This attribute discriminates the different pieces of information based on the level of urgency of their target applications, e.g., data used for safety applications have higher urgency than those required for infotainment.

\begin{table*}[t!]
\caption{Review of the VoI assessment strategies presented in Sec.~\ref{sub:value_evaluator}.}
\label{table:voi_strategies}
\small
\vspace{-0.23cm}
\def\tabularxcolumn#1{m{#1}}
\begin{center}
\renewcommand{\arraystretch}{0.5}
  \begin{tabularx}{\textwidth}{>{\raggedright}>{\hsize=1\hsize\linewidth=\hsize}X>{\hsize=1\hsize\linewidth=\hsize}X}
  \toprule
  Value assessment strategy & Challenges  \\ \midrule
  \emph{Heuristic}: VoI assessment through greedy methods or exhaustive search (it can be considered as a benchmark solution) &
  \begin{itemize}
  	\item Require good empirical functions
	\item Power consumption and runtime limitations \vspace{-0.33cm}
  \end{itemize}\\ \midrule

  \emph{Adaptive}: Refine VoI operations through feedback (via point-to-point transmissions or edge-assisted operations)&
  \begin{itemize}
  	\item Latency (for edge-assisted methods)
  	\item Communication overhead \vspace{-0.33cm}
  \end{itemize}\\ \midrule
  
  	\emph{Machine Learning}:  VoI depends on the correlation (signals can be estimated from a combination of already  available knowledge) &
  	\begin{itemize}
  		\item Require database (not publicly available)
  		\item Computationally inefficient (for limited on-board resources) \vspace{-0.33cm}
  	\end{itemize}\\ \midrule
\emph{Analytic}: VoI assessment through mathematical models (e.g., stochastic theory, information theory, empirical evaluations) & 
\begin{itemize}
	\item Not suitable for real-time VoI estimations
	\item Based on subjective comparative judgments (AHP)\vspace{-0.33cm}
\end{itemize}\\ \bottomrule
  \end{tabularx}
\end{center}
\vspace{-0.5cm}
\label{tab:voI_relwork}
\end{table*}



\paragraph{Generalizability} 
This attribute captures the extent of the interest of the information to multiple applications, e.g., \gls{gps}  can be used by applications ranging from infotainment to safety. Such pieces of information should  be awarded higher value than those of interest for a narrower set of services.

\paragraph{Novelty} 
%
%
The {novelty} attribute captures the relative importance of a certain piece of data with respect to the standard flow generated by a source, e.g., a piece of information that can be easily predicted by the receiver, based on the available knowledge and past observations, will have low novelty. 
 
\paragraph{Provenance (Information Path)} 
This attribute refers to the entire end-to-end source-to-destination path that the  data has followed.
If the message has been relayed through multiple hops (e.g.,  platooned nodes) the carried information  may have been more likely accessed or corrupted by malicious attackers, thereby limiting its integrity and, consequently, its  value.
The \gls{voi} assignment may also be based on the \emph{trust} of the destination towards the source providing the information, which  may result from past interactions between the endpoints. 



\section{How to Assess the Value of Information} 
\label{sec:voi_assignment_techniques}

Although calculating  \gls{voi}  under unpredictable conditions is not an easy task, still, to date, many literature works have proposed  strategies to rate the utility provided by information to specific  applications.
Along these lines, in Sec.~\ref{sub:value_evaluator} we provide a selection of methods that we believe are the most effective to infer  VoI, and in Sec.~\ref{sub:data_scheduler_and_dissemination} we discuss how to efficiently disseminate the most valuable pieces of information over wireless~networks.

\vspace{-0.2cm}
\subsection{Value Assessment} 
\label{sub:value_evaluator}
As illustrated in Fig.~\ref{fig:scheme},
automotive applications annotate each piece of information they generate with \emph{metadata} to facilitate the value assessment operations. Metadata can include (i) type of information, (ii) timestamp of when the information is generated, (iii) network-level requirements to be satisfied,  (iv) importance level of data, (v) accuracy and resolution of data, (vi) source of information.
The application streams are then processed by a \emph{value assessment} block which predicts the utility of each information component for potential receiver(s) under the given context. 
A list  of value assessment schemes includes the following proposals, as summarized in Table~\ref{tab:voI_relwork}.

\paragraph{Heuristic Approaches} 
\label{par:heuristic_approaches}

Heuristic strategies (e.g., greedy methods) are well-known solutions for performing excellent \gls{voi} assessment  when good empirical functions are available for a specific application domain,   and can  therefore be used as a benchmark against other solutions.
However, they may fail  when constrained by runtime limitations or insufficient network resources, and may suffer from significant power consumption and non-negligible delays.

\paragraph{Adaptive Approaches} 
These techniques hierarchically refine value assessment operations by relying on feedback messages which describe how helpful the received information was in relation with the requirements of target applications.
Both distributed and edge-assisted approaches are being discussed, and the trade-off involves signal latency, power consumption,  system overhead, and cost.
In the first case, the endpoints exchange feedback messages through point-to-point transmissions while, in the second case, the feedback is relayed through edge/cloud computing systems.
While incurring communication overhead for both data collection and model distribution, edge-assisted solutions leverage a much larger data set for model training than in a distributed strategy, thereby resulting in a more accurate  \gls{voi} estimation.
In turn, distributed methods can process the feedback in real time, thereby yielding more responsive operations, a critical requirement for most safety-related services.

\paragraph{\gls{ml} Approaches} 
\label{sub:machine_learning_approaches}
Generative Deep Neural Networks can be used to  measure the mutual information of different combinations of the sensory readings and dynamically assign them value scores which depend on the degree of correlation.
Similarly, autoencoders can be trained in an unsupervised manner to  extract features from input vectors and predict the a posteriori probability of a sequence given its entire history.
\gls{ml} can also develop models that link VoI to actions (e.g., settings of link parameters) and effects (e.g., the corresponding performance metrics), in such a way that actions are optimized to the specific operational scenario.
Despite some encouraging features,  \gls{ml} approaches require a large amount of sensory observations for training, which are often not publicly available. 
Furthermore, ML operations  can hardly be completed in low latency, especially considering the limited on-board computational resources of budget car~models.

\paragraph{Analytic Approaches} 
Analytic approaches achieve \gls{voi} estimation through mathematical models.
Stochastic methods, i.e., continuous-time Markov chains, recreate network scenarios and estimate the value that the availability of various sensor information  might bring to the receiver at different times.
Information theory can then be applied to quantify the expected \gls{voi} based on its novelty for potential receivers~\cite{higuchi2019value}.
These approaches provide fine-grained analysis of \gls{voi} tuned to the specifics of the modeled scenario but are not suitable for real-time \gls{voi}.
\gls{ahp} methods~\cite{mu2018understanding}  can be employed to value information in  various application domains based on pairwise comparisons of specific criteria  and to ultimately score the different data dissemination alternatives. 
However, the AHP methodology is not an absolute decision making technique, since relative importance levels are empirically determined based on subjective comparative judgments. For a more detailed description of the \gls{ahp} method we refer to~Sec.~\ref{sec:voi_assessment_the_analytic_hierarchy_process_approach}.

For completeness, we must mention that value assessment operations should be preceded by preliminary on-board processing of sensor observations, to allow the sender to validate the integrity of the acquired information and determine whether it embeds valuable characteristics for receiver(s). This also prevents circulation of redundant or duplicate data~\cite{giordani2019framework}.

\vspace{-0.33cm}
\subsection{Data Scheduler} 
\label{sub:data_scheduler_and_dissemination}

 The \emph{data scheduler} sorts the pieces of information  in a descending order of value and sequentially forwards them to the surrounding receivers. 
The scheduling decision must depend on the type of service that the acquired  measurements intend to provide. 
In particular, a distinction between critical and non-critical services must be made (in the last case, pieces of data  can be aggregated or modified at the application layer before being actually exchanged among the nodes to promote resource consumption minimization).


\vspace{-0.33cm}
\subsection{Dissemination} 
\label{sub:information_dissemination}
Timely and accurate data dissemination  must be guaranteed through any type of wireless  interface for V2X communications, including, but not limited to, IEEE 802.11p, LTE, Wi-Fi and the \gls{mmwave} technology. 
Generally,  IEEE 802.11p and LTE systems  offer relatively low-rate connectivity but guarantee very stable and robust transmissions at short/medium distances (thanks to the intrinsic stability of the low-frequency channels and the omnidirectional transmissions). 
Conversely, \gls{mmwave} systems support very high-throughput connections but exhibit high instability due to the severe signal propagation characteristics and the need to maintain beam alignment~\cite{giordani2018feasibility,giordaniperformance2018}.
In this context, dissemination operations can be improved by using multiple radios in parallel (i.e., \emph{hybrid networking}) to complement the limitations of each type of network.

\vspace{-0.1cm}
\section{VoI Assessment: the AHP Approach} 
\label{sec:voi_assessment_the_analytic_hierarchy_process_approach}




In this section we  consider an illustrative example that explains how to compute  \gls{voi} using the  \gls{ahp} technique~\cite{mu2018understanding} introduced in Sec.~\ref{sub:value_evaluator}. 
We chose the \gls{ahp} strategy for its generality and computational simplicity.
In this example, for the ease of the simulations, we focus  on a subset of the attributes (i.e.,  time/space dependency and information quality), information sources (i.e., surrounding and position information) and applications (i.e.,  safety) described in Sec.~\ref{sec:voi_parameter_preliminaries}.
\medskip

\emph{Step 1: Determination of the attribute weights.} 
First,  \gls{ahp}  derives the relative priorities for~the attributes by populating a pairwise comparison matrix~$M_a$ (represented in Table~\ref{tab:attributes}) with comparison scores (ranging from  1/9 to 9  according to the Saaty comparison scale~\cite{saaty1990decision}) assessing the value of the attributes in the row relative to those in the~column.
For instance, we chose to constrain time-dependency   to be ``moderately more important'' than  information quality (the Saaty score is set to $3$) since, even though the network requires  safety-related information to be reliable, it still needs to prioritize  timely dissemination.
The interdependency of space and quality 
attributes is finally modeled as a function of $\gamma$.

As soon as the comparison scores have been determined, attribute weights $w_a$ are computed~by evaluating the normalized principal eigenvector of $M_a$, i.e., the  eigenvector  that corresponds to the eigenvalue with the largest magnitude.
\medskip

\emph{Step 2: Determination of the conditional information weights.} 
\gls{ahp} is now used for ranking the different kinds of information  to be disseminated.
First, the algorithm compares how each piece of information fares against the others for each of the attributes defined in the previous step.
For example, we assume that, along the quality attribute, the distribution of high-resolution sensory images (which represent surrounding information)  is ``strongly more important'' than the dissemination of precise position information to support safety-related applications.
Second, the conditional weights $w_{d|a}$ for each information source, conditioned to each attribute, are derived by calculating the normalized principal eigenvectors.
\medskip

\begin{table}[t!]
\setcounter{table}{1}
\caption{Pairwise comparison matrix $M_a$ of \gls{voi} attributes for safety applications. Interdependency among attributes is modeled as a function of $\gamma$.} 
\footnotesize
\centering
\begin{tabular}{c|ccc}
\Xhline{2\arrayrulewidth}
\multirow{2}{*}{Attribute}& \multicolumn{3}{c}{Application: Safety}             \\ \cline{2-4} 
& \begin{tabular}[c]{@{}c@{}}Time\\Dependency\end{tabular} & \begin{tabular}[c]{@{}c@{}}Space\\Dependency\end{tabular}        & \begin{tabular}[c]{@{}c@{}}Information\\Quality\end{tabular}      \\\Xhline{2\arrayrulewidth}
\multicolumn{1}{l|}{Time-Dependency}        & 1               & 1        & 3 \\ \cline{2-4} 
\multicolumn{1}{l|}{Space-Dependency}      & 1      & 1                & $\gamma$  \\ \cline{2-4} 
\multicolumn{1}{l|}{Information Quality}               & 1/3       & 1/$\gamma$       & 1       \\ \Xhline{2\arrayrulewidth}
\end{tabular}\vspace{-0.4cm}
\label{tab:attributes}
\end{table}

\begin{figure}[t!]
\centering
    \setlength{\belowcaptionskip}{-0.5cm}
    \includegraphics[width=0.85\columnwidth]{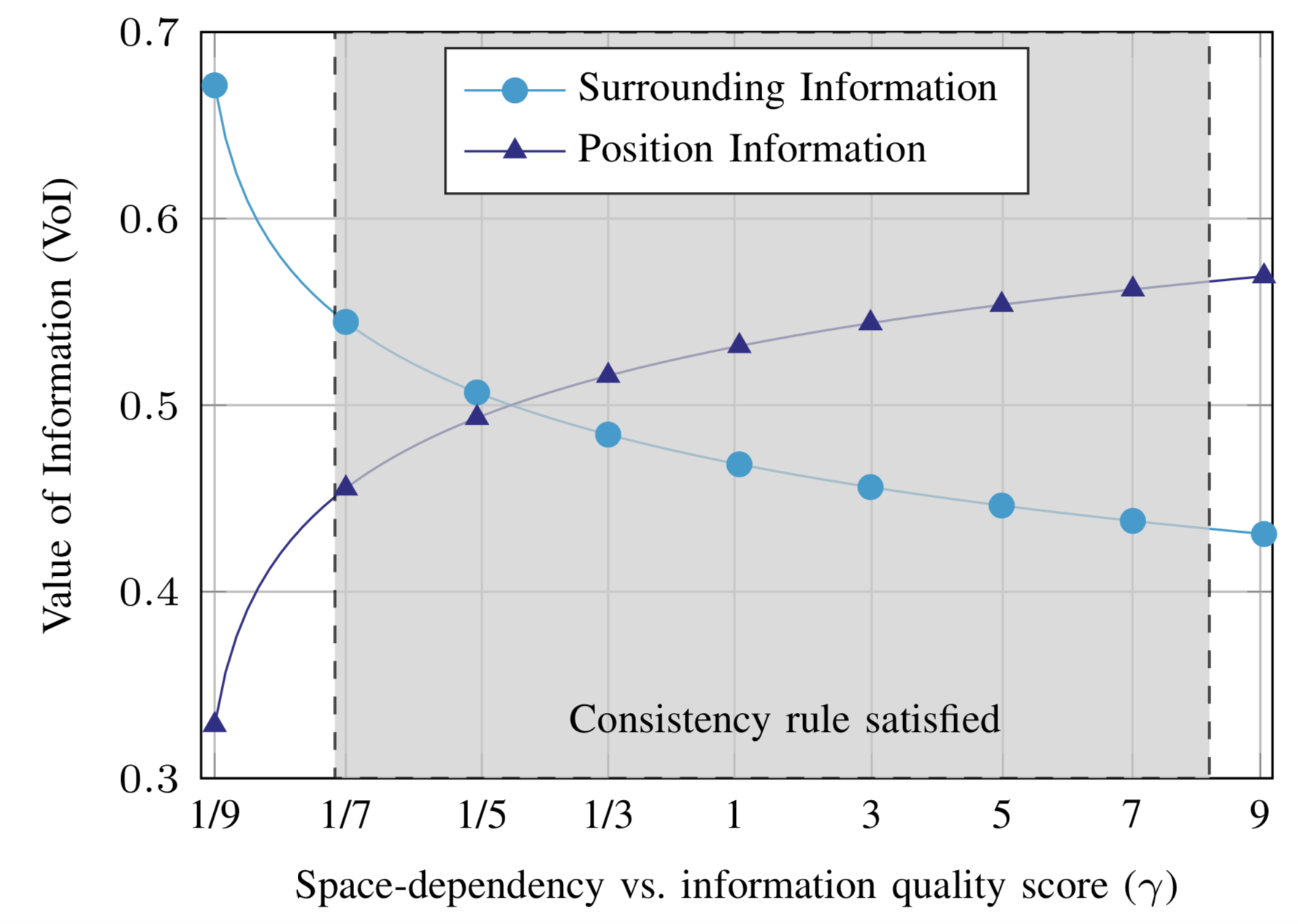}
    \caption{Value of information for various time-dependency vs. information quality scores. Values within the gray area are derived from pairwise comparison scores $\gamma$ that satisfy the consistency rule~\cite{mu2018understanding}.}
    \label{fig:VoI_scores}
\end{figure}

\emph{Step 3: VoI Determination.} 
AHP assigns value to information by multiplying the attribute weights $w_a$ from Step~1  with the conditional information weights $w_{d|a}$ from Step 2.
The result of this procedure is shown in Fig.~\ref{fig:VoI_scores} as a function of~$\gamma$. 
Notice that our conclusions have both mathematical validity, as the comparison scores have been derived from a ratio scale, and an empirical interpretation, since
AHP assumes that the pairwise comparison process involves domain experts that assess which attributes and information sources are more important than others. 
As expected, the value of surrounding and position information goes opposite ways since dissemination of one kind of data must be prioritized at the expense of the other. 
Moreover, we observe that, since in Step 2 we chose to prefer surrounding information over position information along the quality attribute, the overall value of surrounding information drops significantly with the reduction in quality's relative importance as $\gamma$ increases.
This makes intuitive sense as it becomes increasingly less valuable to share inaccurate perception data.
Finally, AHP requires that the attribute comparison scores in $M_a$ satisfy a \emph{consistency rule}, defined in \cite{mu2018understanding}, which imposes that their determination be based on some rationality which guarantees that the assigned attribute interdependencies are fully representative of the application  under consideration.
In Fig.~\ref{fig:VoI_scores}  we have marked in gray the range of $\gamma$ for which the consistency rule is satisfied.

\vspace{-0.33cm}
\section{Conclusions and Open Challenges} 
\label{sec:conclusions_and_open_chall}
Assigning \gls{voi} is fundamental to discriminate the importance of the different information sources, in order to prevent the overload of transmission links.
In this article, we characterized VoI in vehicular networks and investigated data dissemination methods to tackle capacity issues.
We showed that the \gls{voi}  depends on the  environment in which the nodes are deployed and evolves as a function of spatial, temporal and quality~criteria.

This work opens up  some  interesting  research directions.
In particular, intelligent \gls{voi}-aware solutions, able to capture the ever evolving characteristics of the vehicular environment and to dynamically adapt the dissemination scheme accordingly (e.g., based on feedback messages from the receivers or on learning strategies), should be designed.
Moreover, the algorithms should regularly forecast the future \gls{voi} of a given data for all the potentially interested destination nodes based on the knowledge  that has already been acquired through previous  observations.
The investigation of these challenges is still an open issue and will be part of our future research.

\vspace{-0.33cm}
\bibliographystyle{IEEEtran}
\bibliography{bibliography.bib}

\end{document}